\documentstyle[11pt,cite]{article}
\catcode`\@=11
\begin{document}

\begin{titlepage}

\hfill{DFPD/02/TH/29}

\hfill{hep-th/0212253}

\vspace{2 cm}

\centerline{{\huge{Seiberg--Witten Duality}}}
\vspace{0.5 cm}
\centerline{{\huge{in Dijkgraaf--Vafa Theory}}}

\vspace{2 cm}

\centerline{Marco Matone}
\vspace{1 cm}
\centerline{Dipartimento di Fisica ``G.
Galilei''} \centerline{Istituto Nazionale di Fisica Nucleare}
\centerline{Universit\`a di Padova, Via Marzolo, 8 -- 35131
Padova, Italy}

\vspace{2.8cm}

\noindent We show that a suitable rescaling of the matrix model coupling
constant makes manifest the duality group of the $N=2$ SYM theory
with gauge group $SU(2)$. This is done by first identifying the
possible modifications of the SYM moduli preserving the monodromy
group. Then we show that in matrix models there is a simple
rescaling of the pair $(S_D,S)$ which makes them dual variables
with $\Gamma(2)$ monodromy. We then show that, thanks to a crucial
scaling property of the free energy derived perturbatively by
Dijkgraaf, Gukov, Kazakov and Vafa, this redefinition corresponds
to a rescaling of the free energy which in turn fixes the
rescaling of the coupling constant. Next, we show that in terms of
the rescaled free energy one obtains a nonperturbative relation
which is the matrix model counterpart of the relation between the
$u$--modulus and the prepotential of $N=2$ SYM. This suggests
considering a dual formulation of the matrix model in which the
expansion of the prepotential in the strong coupling region, whose
QFT derivation is still unknown, should follow from perturbation
theory. The investigation concerns the $SU(2)$ gauge group and can
be generalized to higher rank groups.

\end{titlepage}

\newpage

\noindent Recently Dijkgraaf and Vafa derived crucial relations
between matrix models and SYM theories \cite{1,2,3}. Subsequently,
in \cite{DGKV} Dijkgraaf, Gukov, Kazakov and Vafa provided the
explicit relationship between the $N=2$ SYM theory \cite{SW} and
matrix models. The original proposal was based on geometrical
engineering analysis in string theory, while in \cite{n,n+1} it
has been argued that there exists a QFT proof of the
Dijkgraaf--Vafa formulation. In these derivations a crucial role
is played by holomorphy. This is a crucial issue as, for example,
holomorphy and symmetries are at the basis of $N=2$ SYM duality.
Therefore, a basic question in considering the matrix model
formulation, is to identify the duality structure which is the
essence of Seiberg--Witten theory \cite{SW}. There are several
reasons which suggest introducing the powerful tool of duality
directly in the matrix model formulation. For example, an
interesting question would be to understand the analogous of the
nonperturbative relation between the $u$--modulus and the
prepotential \cite{rela}. This should be useful for a proof of the
relationship between matrix model and $N=2$ SYM along the lines of
\cite{proof}. We also note that this relation, which has been
useful in investigating related issues \cite{rela2}, should help
in deriving possible exact results in matrix models. Furthermore,
this duality may help in understanding what is the QFT formulation
of $N=2$ SYM in the strong coupling region. In this context, one
should expect that the expansion of the $N=2$ SYM prepotential in
the strong coupling region should be obtained by means of a
perturbative calculation in a dual matrix model formulation.

The aim of this paper is to introduce such a duality in matrix
models. We will start by showing that, on general grounds, in
order to preserve the Seiberg--Witten duality, only a class of
redefinitions of the moduli $(a_D,a)$ is allowed. This is based on
a mathematical general observation which involves the
Picard--Fuchs equation.\footnote{A suitable generalization of the
method introduced here, suggests a possible application in
investigating the Picard--Fuchs equations in the framework of the
Mirror conjecture.} In particular, it is shown that if
$$
\tau(a)={\partial{\cal S}_D\over\partial{\cal S}}={\partial
a_D\over\partial a},
$$
with $\tau$ the $N=2$ effective coupling constant, then $({\cal
S}_D,{\cal S})$ have the same monodromy of $(a_D,a)$ on the
$u$--plane if
\begin{equation}
{\cal S}_D=fa_D+4(u^2-\Lambda_{SW}^4)f'a_D' ,\qquad {\cal
S}=fa+4(u^2-\Lambda_{SW}^4)f'a', \label{introsx}\end{equation}
with $f$ an arbitrary singlevalued function of $u$ (note that a
possible additional $Z_2$ monodromy leaves $\tau$ invariant).

Next, we identify the explicit relationship between the matrix
model variables $(S_D,S)$ and $(a_D,a)$. It turns out that
$(S_D,S)$ cannot have $\Gamma(2)$--monodromy. Nevertheless,
remarkably, the simple rescaling
$$
{\cal S}=\left({\Lambda_{SW}\over 2^{3/2} u^{1/2}}\right)^3S,
$$
restores duality, that is $${\cal S}={\Lambda_{SW}^{3}\over 3\cdot
2^6}\left[ u^{-1/2}a-2(u^2-\Lambda_{SW}^4)u^{-3/2}a'\right],$$
satisfies (\ref{introsx}) with $f={\Lambda_{SW}^3\over \sqrt 2
\cdot 48} u^{-1/2}$. On the other hand, this fixes $S^D$ to be
$$
{\cal S}_D={\Lambda_{SW}^3\over3\cdot 2^6}\left[
u^{-1/2}a_D-2(u^2-\Lambda_{SW}^4)u^{-3/2}a_D'\right], $$ which, in
turn, defines ${\cal F}_0$ by
$$
{\cal S}_D={\partial {\cal F}_0\over \partial {\cal S}}. $$ It
then follows that the new pair has $\Gamma(2)$ monodromy
\begin{equation}
\left(\begin{array}{c} \tilde {\cal S}_D\\ \tilde {\cal S}
\end{array}\right)= \left(\begin{array}{c}A\\C
\end{array}\begin{array}{cc}B\\D\end{array}\right)\left(\begin{array}{c}
{\cal S}_D\\ {\cal S}
\end{array}\right).
\label{2x}\end{equation} We then show that thanks to a remarkable
scaling property of the genus zero free energy ${\cal F}_{0}(
S_k,\Delta,\Lambda)$, passing to the new variables is equivalent
to a simple rescaling, that is the change of variables
($\Lambda=2^{-1/2}\Lambda_{SW}$)
$$
S_k\longrightarrow {\cal S}_k=\left({\Lambda\over \Delta}\right)^3
S_k, \qquad \Delta\longrightarrow {\Lambda\over \Delta}
\Delta=\Lambda, \qquad \Lambda\longrightarrow {\Lambda\over
\Delta}\Lambda= {\Lambda^2\over \Delta},
$$
induces the scaling transformation
$$
{\cal F}_{0}\longrightarrow {\cal F}_0\left({\cal
S}_k,\Lambda,{\Lambda^2\over\Delta}\right)=
\left({\Lambda\over\Delta}\right)^6{\cal F}_{0}(
S_k,\Delta,\Lambda),
$$
which has no effect in deriving the critical values since the
factor $\left({\Lambda\over\Delta}\right)^6$ cancels the one from
the Jacobian in $\tau_{ij}$, that is $\partial^2\mu^6{\cal
F}_{0}/\partial {\cal S}_i\partial {\cal S}_j=\partial^2{\cal
F}_{0}/\partial S_i\partial S_j$. As a result, even if the
partition function remains invariant, we have the same rescaling
for both the potential and the matrix coupling constant
$$
g_S\longrightarrow g_{\cal S}=\left({\Lambda\over\Delta}\right)^3
g_S,$$ $$ W\longrightarrow{\cal
W}(\Phi)=\left({\Lambda\over\Delta}\right)^3 W(\Phi).$$ As a
consequence the scaling generalizes to
$$
{\cal F}_{g}\longrightarrow {\cal F}_g\left({\cal
S}_k,\Lambda,{\Lambda^2\over\Delta}\right)=
\left({\Lambda\over\Delta}\right)^{3(2-2g)}
{\cal F}_{g}( S_k,\Delta,\Lambda).
$$
We then show that the new prepotential, which is obtained by
integrating with respect to ${\cal S}$, the function $\tau$ at the
extremum, satisfies the nonperturbative relation
$$
\left({\Lambda\over\Delta}\right)^4={48\pi
i\over\Lambda^6}\left({\cal F}_0-{{\cal S}\over 2}{\partial {\cal
F}_0\over\partial {\cal S}}\right).
$$
Introducing duality then leads to consider a dual formulation of
the matrix model that we propose should correspond to introduce
the Legendre transform of the free energy $$ {\cal F}_{Dg}={\cal
F}_g-\sum_{i=1,2}{\cal S}_i{\partial {\cal F}_g\over\partial {\cal
S}_i},$$ where now ${\cal F}_g\equiv{\cal F}_g\left({\cal
S}_k,\Lambda,{\Lambda^2\over\Delta}\right)$.

Let us start by recalling that in matrix model the effective
coupling constant of $N=2$ SYM has the form \cite{DGKV}
\begin{equation}
\tau(a)={\partial^2 {\cal F}_0(S)\over \partial S^2},
\label{1}\end{equation} which should be compared with
\begin{equation}
\tau(a)={\partial^2 {\cal F}(a)\over \partial a^2}.
\label{2}\end{equation} The problem is to find the relationship
between ${\cal F}_0(S)$ and ${\cal F}(a)$. Let us introduce the
dual
\begin{equation}
S_D={\partial {\cal F}_0(S)\over \partial S},
\label{3}\end{equation} so that
\begin{equation}
\tau(a)={\partial_u S_D\over \partial_u S}={\partial_u a_D\over
\partial_u a}. \label{5}\end{equation}
It is clear that the dual pairs $(S_D,S)$ and $(a_D,a)$ should
have the same monodromy on the $u$--plane. As observed in
\cite{BIM} in considering a similar problem, we may use the
differential equation \cite{EQ,rela}
\begin{equation}
\left(\partial_u^2+\frac{1}{4(u^2-\Lambda_{SW}^4)}\right)
\pmatrix{a_D\cr a}=0, \label{embhe}\end{equation} to investigate
the structure of the possible solutions of (\ref{5}). Generalizing
the analysis in \cite{BIM} we set\footnote{We are using the
notation $({{\cal{S}}_D},{\cal S})$ rather than $(S_D,S)$ since,
as we will see, the pair $(S_D,S)$ defined in matrix model has not
$\Gamma(2)$--monodromy.} ($'\equiv\partial_u$)
\begin{equation}
{\cal S}_D=f_Da_D+g_Da_D' ,\qquad {\cal S}=fa+ga' ,
\label{wodihw}\end{equation} where the two dual pairs $(f_D,f)$
and $(g_D,g)$ are functions of $u$. Note that if these functions
are singlevalued with respect to $u$, then $({\cal S}_D,{\cal S})$
would have the $\Gamma(2)$ monodromy of $(a_D,a)$. However, since
a possible additional $Z_2$ monodromy of $({\cal S}_D,{\cal S})$
with respect to $(a_D,a)$ does not change the polymorphicity of
${\cal S}_D'/{\cal S}'$, the functions $(f_D,f)$ and $(g_D,g)$
should be singlevalued on the $u$--space except for a possible
minus sign they may get winding around some point.

We now show that if the functions $(f_D,f)$ and $(g_D,g)$ solve a
differential equation, then Eq.(\ref{5}) is satisfied. By
(\ref{embhe}) and (\ref{wodihw}) we have
$$
{\cal S}_D'=\tilde f_Da_D+\tilde g_Da_D',\qquad {\cal S}'= \tilde
fa+\tilde ga',
$$ where
$$
\tilde f_D=f_D'-\frac{1}{4(u^2-\Lambda_{SW}^4)}g_D, \qquad \tilde
g_D=f_D+g_D',
$$
$$
\tilde f=f'-\frac{1}{4(u^2-\Lambda_{SW}^4)}g, \qquad \tilde
g=f+g'.$$ Imposing $\tilde f_D=0=\tilde f$
\begin{equation}
g_D=4(u^2-\Lambda_{SW}^4)f_D',\qquad g=4(u^2-\Lambda_{SW}^4)f',
\label{eiuoh}\end{equation} and $\tilde g_D=\tilde g$, that is
\begin{equation}
\tilde g_D=f_D+8uf_D'+4(u^2-\Lambda_{SW}^4)f_D''=
f+8uf'+4(u^2-\Lambda_{SW}^4)f''=\tilde g,
\label{oiuce2}\end{equation} we obtain
$$
{\cal S}_D'=h a_D',\qquad {\cal S}'=h a',$$ where $h\equiv\tilde
g_D=\tilde g$. Since $f_D$ and $f$ satisfy the same differential
equation (\ref{oiuce2}), it follows that once either $f_D$ or $f$
is given, say $f$, besides the choice $f_D=f$ (which would imply
$g_D=g$), one can also choose $f_D$ to be any other solution of
(\ref{oiuce2}). Summarizing, from (\ref{wodihw}) and (\ref{eiuoh})
we have
\begin{equation}
{\cal S}_D=f_Da_D+4(u^2-\Lambda_{SW}^4)f_D'a_D' ,\qquad {\cal
S}=fa+4(u^2-\Lambda_{SW}^4)f'a', \label{wodihwn}\end{equation} and
${\cal S}_D'/{\cal S}'=a_D'/a' =\tau$.

Let us start considering the relationship between the $N=2$ SYM
and matrix model variables. We first set
$$\Lambda=2^{-1/2}\Lambda_{SW},\qquad \Delta^2=4u,
$$ in the loop expansion of $S$ \cite{DGKV}
\begin{equation}
{S\over 2^3 u^{3/2}}={1\over 2^6}\left({\Lambda_{SW}^2\over u
}\right)^2+ {3\over2^{11}} \left({\Lambda_{SW}^2\over
u}\right)^4+{35\over 2^{16}} \left({\Lambda_{SW}^2\over u
}\right)^6+\ldots . \label{laserie}\end{equation} We now show that
rather than $S$ itself, it is the right hand side of
(\ref{laserie}) that matches with the expansion of ${\cal S}$ in
(\ref{wodihwn}) with
\begin{equation}
f={1\over \sqrt 2 \cdot 48} u^{-1/2}.
\label{effediu}\end{equation} Therefore, while ${S\over 2^3
u^{3/2}}$ is of the form that preserves duality, this is not the
case for $S$ itself. As we will see, this will lead to a natural
rescaling of the matrix model coupling constant which will make
Seiberg--Witten duality manifest. In particular, we will see that
one has to rescale $S$ to
\begin{equation}
\left({\Lambda_{SW}\over 2^{3/2}u^{1/2}}\right)^3S
={\Lambda_{SW}^3u^{-3/2}\over3\cdot 2^6}\left[
ua-2(u^2-\Lambda_{SW}^4)a'\right].
\label{almostesse}\end{equation} In order to compare
(\ref{laserie}) and (\ref{almostesse}) we expand $a$ for
$u\to\infty$
$$
a(u)={\sqrt 2\over \pi}\int_{-\Lambda_{SW}^2}^{\Lambda_{SW}^2}
dx{\sqrt{x-u}\over \sqrt{x^2-\Lambda_{SW}^4}}
$$
$$
=\sqrt {2u}\left(1- {1\over 2^4}\left({\Lambda_{SW}^2\over
u}\right)^2-{15\over2^{10}} \left({\Lambda_{SW}^2\over
u}\right)^4-{105\over2^{14}} \left({\Lambda_{SW}^2\over
u}\right)^6 +\ldots \right), $$ that substituted in
(\ref{almostesse}) exactly reproduces (\ref{laserie}).
Substituting  (\ref{almostesse}) in (\ref{5}) and using
\begin{equation} S'={1\over \sqrt2 \cdot
4}(a-2ua'),\label{Sprimo}\end{equation} we see that the relation
between $(S_D',S')$ and $(a_D',a')$ is rather involved
$$
S_D'={1\over \sqrt 2\cdot 4}(a-2ua'){a_D'\over a'}.
$$
This is not only a formal question since $S_D'$ and $S'$ cannot
have simultaneously $\Gamma(2)$ monodromy. Even if this is
implicit in the above construction, it is instructive to
illustrate it explicitly. In particular, if $S$ has $\Gamma(2)$
monodromy, this cannot be the case for $S_D$. Since the monodromy
commutes with the derivative, we show this for $S'_D$ and $S'$.
Under the action of $\Gamma(2)$ we have
$$
S'\longrightarrow \gamma(S')={1\over \sqrt 2\cdot 4}C(a_D-2ua_D')+
{1\over \sqrt 2\cdot 4}D(a-2ua'),
$$
so $S'$ has $\Gamma(2)$ monodromy iff we consider as its dual
$$
\hat S_D'={1\over \sqrt 2\cdot 4}(a_D-2ua_D')\neq S_D',
$$
so that
$$
\gamma(S')=C \hat S_D'+D S'.
$$
Of course, as follows by the previous analysis, even if $\hat
S_D'$ and $S'$ have $\Gamma(2)$ monodromy, their ratio cannot
correspond to $\tau$.

A similar reasoning holds for $S_D'$. Actually, since under
$\Gamma(2)$
$$
{A\tau+B\over C\tau+D}={AS_D'+BS'\over
CS_D'+DS'}={\gamma(S_D')\over \gamma(S')},
$$
we see that
$$
\gamma(S'_D)={(AS_D'+BS')(C\hat S_D'+DS')\over CS_D'+DS'},
$$
which cannot correspond to the $\Gamma(2)$ monodromy, that is
$$
\gamma(S_D')\neq AS'_D+BS'.
$$
Note that
\begin{equation}
S_D={1\over \sqrt2\cdot 2^2}\int^u_{u_0} d\tilde u (\tau a
-2\tilde u\partial_{\tilde u}a_D)+S_D(u_0),\qquad S= {1\over
\sqrt2\cdot 6} (ua-2(u^2-\Lambda_{SW}^4)a'),
\label{oiiuxh}\end{equation} and by (\ref{almostesse}) and
(\ref{Sprimo})
$$a={\sqrt2\cdot 2\over\Lambda_{SW}^4}
[3uS-2(u^2-\Lambda_{SW}^4)S'].$$ The fact that the Seiberg--Witten
duality is not manifest with the pair $(S_D,S)$ can be also seen
by noticing that $S$ solves the differential equation
\begin{equation}
\left(\partial_u^2-{3\over 4(u^2-\Lambda^4_{SW})}\right)S=0,
\label{odcwo}\end{equation} which is not satisfied by $S_D$,
indicating once again that they cannot have the same monodromy on
the $u$--plane. Inverting Eq.(\ref{odcwo}) we obtain
\begin{equation}
4\left({\cal G}^2-\Lambda_{SW}^4\right){\partial^2 {\cal G}\over
\partial S^2}+3{\cal S}\left({\partial {\cal G}\over\partial
S}\right)^3=0, \label{oidhw}\end{equation} where
$$
u={\cal G}(S).
$$

To select a dual pair with $\Gamma(2)$ monodromy and whose ratio
corresponds to $\tau$ is essential to recognize the underlying
geometry of $N=2$ SYM. In particular, winding around the
$u$--moduli space, the pair $(S_D,S)$ will not preserve the
analogous relations satisfied by $(a_D,a)$. In order to restore
manifest duality we rescale $S$ and define
\begin{equation}
{\cal S}=\left({\Lambda_{SW}\over 2^{3/2} u^{1/2}}\right)^3S,
\label{riscalato}\end{equation} that is
\begin{equation}
{\cal S}={\Lambda_{SW}^{3}\over3\cdot 2^6}\left[
u^{-1/2}a-2(u^2-\Lambda_{SW}^4)u^{-3/2}a'\right],
\label{essecorsiva}\end{equation} where the term $\Lambda_{SW}^3$
has been introduced to make $S$ and ${\cal S}$ of the same
dimension. We now choose $f_D=f$, so that by (\ref{wodihwn})
\begin{equation}
{\cal S}_D={\Lambda_{SW}^3\over3\cdot 2^6}\left[
u^{-1/2}a_D-2(u^2-\Lambda_{SW}^4)u^{-3/2}a_D'\right],
\label{essecorsivaduale}\end{equation} which, in turn, defines
${\cal F}_0$ by
$$
{\cal S}_D={\partial {\cal F}_0\over \partial {\cal S}}. $$ By
construction the pair $({\cal S}_D,{\cal S})$ has the same
monodromy of $(a_D,a)$ on the $u$--plane except for a minus sign
they get winding around $u=0$, as observed this does not change
the polymorphicity properties of $\tau$.

We can now use the method introduced in \cite{rela} to derive the
exact relation between the prepotential and the modular invariant.
In this case, by means of $({\cal S}_D,{\cal S})$ we may construct
the modular invariant
\begin{equation}
v={2^{13}\cdot 3\pi i\over\Lambda_{SW}^6}\left({\cal F}_0-{{\cal
S}\over 2}{\partial {\cal F}_0\over\partial {\cal S}}\right),
\label{v}\end{equation} which implies that the pair $({\cal
S}_D,{\cal S})$ satisfies the differential equation
\begin{equation}
\left(\partial_v^2+\frac{1}{2}\{\sigma,v\}\right) \pmatrix{{\cal
S}_D\cr {\cal S}}=0, \label{embhe2}\end{equation} where
$\{g(x),x\}$ denotes the Schwarzian derivative
$g'''/g'-{3\over2}(g''/g')^2$ and $\sigma$ is an arbitrary
M\"obius transformation of the ratio ${\cal S}_D/{\cal S}$. Later
we will see that a simple redefinition of the matrix model
coupling constant precisely leads to the above duality structure.
Furthermore, we will see that $v=\Lambda_{SW}^4/u^2$ and will find
the explicit expression of ${\cal S}(v)$ and ${\cal S}_D(v)$.

We now show that thanks to a scaling property of ${\cal F}_0$, it
is possible to identify the right variables to make
Seiberg--Witten duality in Dijkgraaf--Vafa theory manifest. First,
we note that, by an overall rescaling, the loop expansion of the
genus zero free energy in matrix model \cite{DGKV} reduces by one
the number of variables ($\Lambda=2^{-1/2}\Lambda_{SW}$)
$$\Delta^{-6}
{\cal F}_{0}(S_k,\Delta,\Lambda)=$$
\begin{equation}{1 \over 2} \sum_{i=1,2}
\left({S_i\over\Delta^3}\right)^2\left[\log \left( {S_i \over
\Delta^3} \right)-{3\over2}\right]-\left({S_1\over\Delta^3}
+{S_2\over\Delta^3}\right)^2 \log \left( {\Lambda \over \Delta}
\right)+\sum_{n\geq3}\sum_{i=0}^n c_{n,i}
\left({S_1\over\Delta^3}\right)^{n-i}\left({S_2\over\Delta^3}\right)^i,
\label{dacompararesotto}\end{equation} where
\begin{equation}
c_{n,i}=(-1)^nc_{n,n-i}, \qquad c_{n,i}=(-1)^i|c_{n,i}|,
\label{ojh}\end{equation} so that, except for the first term,
${\cal F}_0$ is symmetric in $S_1$ and $-S_2$. By Euler theorem we
have
\begin{equation}
\sum_{i=1,2}S_i{\partial {\cal F}_0\over\partial
S_i}+{\Delta\over3}{\partial {\cal F}_0\over\partial\Delta}+
{\Lambda\over3}{\partial {\cal F}_0\over\partial\Lambda}=2{\cal
F}_0. \label{opicjas}\end{equation} Eq.(\ref{dacompararesotto})
would suggest that the natural variables are $S_k/\Delta^3$ rather
than $S_k$. However, note that this would change the dimensional
properties, so we should select $\Lambda^3S_k/\Delta^3$.
Furthermore, we should also choose the scale $\Lambda$ as
independent variable. So we should express ${\cal F}_0$ as a
function of
$$
{\cal S}_k=\left({\Lambda\over\Delta}\right)^3 S_k,\qquad
\mu=?,\qquad \Lambda.
$$
It remains to find $\mu$ which, of course, should depend on
$\Delta$ and possibly on $\Lambda$. A closer look to
(\ref{dacompararesotto}) fixes it. Actually,
Eq.(\ref{dacompararesotto}) suggests considering a natural
rescaling of all dimensional quantities of the arguments of ${\cal
F}_0$, by the dimensionless factor ${\Lambda\over \Delta}$. In
particular, if $[x]=[\Lambda]^n$, then $x\to\left({\Lambda\over
\Delta}\right)^n x$, that is
$$
S_k\longrightarrow \left({\Lambda\over \Delta}\right)^3 S_k,
\qquad \Delta\longrightarrow {\Lambda\over \Delta} \Delta=\Lambda,
\qquad \Lambda\longrightarrow {\Lambda\over \Delta}\Lambda=
{\Lambda^2\over \Delta},
$$
and the map we define is
$$
{\cal F}_{0}(S_k,\Delta,\Lambda)\qquad\longrightarrow \qquad{\cal
F}_{0}\left(\left({\Lambda\over \Delta}\right)^3
S_k,\Lambda,{\Lambda^2\over \Delta}\right),
$$
showing that $S_k$, $\Delta$ and $\Lambda$ combine in such a way
that the natural variables for ${\cal F}_0$ are
$$
{\cal S}_1=\left({\Lambda\over \Delta}\right)^3 S_1,\qquad {\cal
S}_2= \left({\Lambda\over \Delta}\right)^3 S_2, \qquad
\mu={\Lambda\over \Delta},\qquad \Lambda.
$$
This also follows by the scaling law which is crucial for us
\begin{equation}
{\cal F}_{0}(\mu^3S_k,\mu\Delta,\mu\Lambda)=\mu^6{\cal
F}_{0}(S_k,\Delta,\Lambda), \label{preoknco}\end{equation} that we
rewrite as
\begin{equation}
{\cal F}_{0}({\cal S}_k,\Lambda,\mu\Lambda)=\mu^6{\cal
F}_{0}(S_k,\Delta,\Lambda). \label{oknco}\end{equation} Note that
$$
{\cal F}_{0}({\cal S}_k,\Lambda,\mu\Lambda)= $$
\begin{equation}\Lambda^6\left\{{1 \over 2} \sum_{i=1,2} \left({{\cal
S}_i\over\Lambda^3}\right)^2 \left[\log \left( {{\cal S}_i \over
\Lambda^3}\right)-{3\over2}\right]-\left({{\cal
S}_1\over\Lambda^3}+{{\cal
S}_2\over\Lambda^3}\right)^2\log\mu+\sum_{n\geq3}\sum_{i=0}^n
c_{n,i} \left({{\cal S}_1\over\Lambda^3}\right)^{n-i}\left({{\cal
S}_2\over\Lambda^3}\right)^i\right\},
\label{dacompararesopra}\end{equation} that differs from ${\cal
F}_{0}({\cal S}_k,\Lambda,\Delta)$, which, we stress, is the
original function with $\Lambda$ and $\Delta$ interchanged and
$S_k$ replaced by ${\cal S}_k$, by the sign of the term $({\cal
S}_1+{\cal S}_2)^2 \log\mu$.

Since ${\cal F}_{0}({\cal S}_k,\Lambda,\mu\Lambda)$ is a function
of ${\cal S}_k$, $\mu$, and $\Lambda$, it follows by (\ref{oknco})
that this is the case also for ${\cal F}_{0}(S_k,\Delta,\Lambda)$.
Therefore, we consider the map $(S_k,\Delta,\Lambda)
\longrightarrow ({\cal S}_k,\mu,\Lambda)$, as change of variables
for ${\cal F}_0(S_k,\Delta,\Lambda)$. The relationships between
the derivatives in the old and new variables are

\begin{equation}
{\partial{\cal F}_{0}\over\partial S_1}=\mu^3{\partial{\cal
F}_{0}\over\partial {\cal S}_1},\qquad {\partial{\cal
F}_{0}\over\partial S_2}=\mu^3{\partial{\cal F}_{0}\over\partial
{\cal S}_2}, \label{d1}\end{equation}

\begin{equation}
{\partial{\cal F}_{0}\over\partial
\Delta}=-3{\mu\over\Lambda}{\cal S}_1 {\partial{\cal
F}_{0}\over\partial {\cal S}_1}-3{\mu\over\Lambda}{\cal
S}_2{\partial{\cal F}_{0}\over\partial {\cal
S}_2}-{\mu^2\over\Lambda}{\partial{\cal F}_{0}\over\partial \mu},
\label{d2}\end{equation}

\begin{equation}
{\partial{\cal F}_{0}\over\partial \Lambda}= {\partial{\cal
F}_{0}\over\partial \Lambda}+3{{\cal
S}_1\over\Lambda}{\partial{\cal F}_{0}\over\partial {\cal
S}_1}+3{{\cal S}_2\over\Lambda}{\partial{\cal F}_{0}\over\partial
{\cal S}_2}+{\mu\over\Lambda}{\partial{\cal F}_{0}\over\partial
\mu}, \label{d3}\end{equation}
\\
\noindent where in the left hand side the derivatives have been
taken considering ${\cal F}_{0}$ as function of the old variables,
while on the right hand side it is seen as function of $({\cal
S}_k,\mu,\Lambda)$. In the following we make an abuse of notation
and drop a factor $\Lambda$, that is
\begin{equation} {\cal F}_{0}({\cal S}_k,\Lambda,\mu)\equiv{\cal F}_{0}({\cal S}_k,
\Lambda,\mu\Lambda)=\mu^6{\cal F}_{0}(S_k,\Delta,\Lambda).
\label{okncopokerdassi}\end{equation} Minimizing
$$
W_{eff}=\sum_{i=1,2}{\partial{\cal F}_0\over\partial S_i},$$ we
obtain, by (\ref{d1}) and (\ref{okncopokerdassi})
$$
\sum_{i=1,2}{\partial^2{\cal F}_0\over
\partial S_i\partial S_j}=
\sum_{i=1,2}\mu^6{\partial^2{\cal F}_0\over \partial {\cal
S}_i\partial {\cal S}_j}=\sum_{i=1,2}{\partial^2{\cal F}_0({\cal
S}_k,\Lambda,\mu)\over
\partial {\cal S}_i\partial {\cal S}_j}=0,
$$
which gives ${\cal S}={\cal S}_1=-{\cal S}_2$, where \cite{DGKV}
$$
{\cal S} =\Lambda^3(\mu^4
+6\mu^8+140\mu^{12}+4620\mu^{16}+\ldots).
$$
The effective coupling constant of $N=2$ SYM with gauge group
$SU(2)$ is given by
$$
\tau={\partial^2 {\cal F}_0\over
\partial S_1\partial S_2}\bigg|_{S_1=-S_2=S},
$$
and by (\ref{d1}) and (\ref{okncopokerdassi})
$$
\tau={\partial^2 {\cal F}_{0}({\cal S}_k,\Lambda,\mu)\over
\partial {\cal S}_1\partial {\cal S}_2}\bigg|_{{\cal S}_1=-{\cal
S}_2={\cal S}},
$$
where here ${\cal F}_0$ is rescaled by $1/\pi i$ with respect to
the one in (\ref{dacompararesotto}). So, we have seen that, thanks
to the scaling property (\ref{okncopokerdassi}), one obtains the
same effective coupling constant $\tau(a)$, if in the matrix model
one considers as variables the old ones rescaled by
$\mu^n=(\Lambda/\Delta)^n$, with $n$ defined by $[x]=[\Lambda]^n$.
As a consequence the duality structure of $N=2$ SYM with gauge
group $SU(2)$ is manifest. Before showing this explicitly we
explain how the above rescaling of ${\cal F}_0$ simply amounts to
a different choice of the matrix model coupling constant. Let us
set
\begin{equation}
g_{\cal S}=\mu^3 g_S, \qquad {\cal
W}(\Phi)=\mu^3W(\Phi),\label{newgS}\end{equation} and note that
\begin{equation}
Z={1\over {\rm vol(G)}}\int d\Phi \exp \left(-{1\over g_S} {\rm
tr}\, W(\Phi)\right)={1\over {\rm vol(G)}}\int d\Phi \exp
\left(-{1\over g_{\cal S}} {\rm tr}\,  {\cal W}(\Phi)\right),
\label{Z}\end{equation} so that
\begin{equation}
Z=\exp\left(-\sum_{g\geq0}g_S^{2g-2}{\cal F}_g\right)=
\exp\left(-\sum_{g\geq0}g_{\cal S}^{2g-2}\tilde{\cal F}_g\right),
\label{expann}\end{equation} where
\begin{equation}
\tilde{\cal F}_g=\mu^{3(2-2g)}{\cal F}_g.
\label{iuc}\end{equation} In particular, by
(\ref{okncopokerdassi}) we see that $\tilde{\cal F}_0=\mu^6{\cal
F}_0={\cal F}_0({\cal S}_k,\Lambda,\mu)$. This indicates that also
the higher genus contributions should be considered as functions
of the new variables, that is
\begin{equation}
\tilde{\cal F}_g=\mu^{3(2-2g)}{\cal F}_g={\cal F}_g({\cal
S}_k,\Lambda,\mu), \label{scalareale}\end{equation} so we rewrite
\begin{equation}
Z=\exp\left(-\sum_{g\geq0}g_{\cal S}^{2g-2}{\cal F}_g\right),
\label{expannoliik}\end{equation} where now ${\cal F}_g\equiv{\cal
F}_g({\cal S}_k,\Lambda,\mu)$.

Let us now derive the explicit expression for ${\cal S}_D$ and
${\cal S}$ and show how the rescaling leads to make the $N=2$ SYM
duality manifest. The trick is to first consider the derivative of
$v$ with respect to $u$. In particular, by (\ref{essecorsiva}) and
(\ref{essecorsivaduale}) we have
\begin{equation}
{\cal S}_D'=-{\Lambda_{SW}^7\over64} u^{-5/2}a_D',\qquad {\cal
S}'=-{\Lambda_{SW}^7\over64} u^{-5/2}a',
\label{iprimati}\end{equation} and by (\ref{v})
\begin{equation}
v'={2^{10}\cdot3{\pi i}\over\Lambda_{SW}^6}({\cal S}_D{\cal
S}'-{\cal S}{\cal S}_D')=\pi i\Lambda_{SW}^4(a_D'a-a_Da')u^{-3}.
\label{oiajx}\end{equation} On the other hand, since
$aa_D'-a_Da'=2i/\pi$, we have
\begin{equation}
v'=-2{\Lambda_{SW}^4} u^{-3}, \label{oiajx2}\end{equation} that is
\begin{equation}
v= \left({\Lambda_{SW}^2\over u}\right)^2,
\label{oiajx3}\end{equation} where the additive constant, that
corresponds to fix the additive constant of ${\cal F}_0$, has been
set to zero. By construction we know that ${\cal S}$ satisfies a
second order differential equation with respect to $v$ in which
the first derivative term is absent. Actually, taking the second
derivative of ${\cal S}$ with respect to $v$, we have
\begin{equation}
\partial^2_v {\cal S}=-(\partial_u
v)^{-3}\partial^2_uv\partial_u{\cal
S}+(\partial_uv)^{-2}\partial_u^2{\cal S}={3u^4\over
16\Lambda_{SW}^4(u^2-\Lambda_{SW}^4)} {\cal S},
\label{vabene}\end{equation} that is $({\cal S}_D,{\cal S})$
satisfy the second order differential equation
\begin{equation}
\left(\partial_v^2+{3\over 16v(v-1)} \right) \pmatrix{{\cal
S}_D\cr {\cal S}}=0, \label{embhe23}\end{equation} whose solutions
can be obtained directly by (\ref{essecorsiva}) and
(\ref{essecorsivaduale}) using $a_D(u(v))$ and $a(u(v))$
\begin{equation}
{\cal S}_D={\Lambda_{SW}^3\sqrt v\over\sqrt2\cdot96
\pi}\int_{-1}^{1\over\sqrt v} dx {x-\sqrt v\over
\sqrt{x^2-1}\sqrt{\sqrt v x -1}},\qquad {\cal
S}={\Lambda_{SW}^3\sqrt v\over\sqrt2\cdot96 \pi}\int_{-1}^1 dx
{x-\sqrt v\over \sqrt{x^2-1}\sqrt{\sqrt v x -1}}.
\label{opidjc}\end{equation} Inverting Eq.(\ref{embhe23}) we
obtain the differential equation for $v={\cal H}({\cal S})$
\begin{equation}
16{\cal H}(1-{\cal H}){\partial^2 {\cal H}\over\partial {\cal
S}^2}+3{\cal S}\left({\partial {\cal H}\over\partial {\cal
S}}\right)^3=0. \label{iowuxch}\end{equation} Since
$$
\mu=\left({\Lambda_{SW}^2\over 2^3 u}\right)^{1/2}\longrightarrow
v=2^6\mu^4,
$$
we have
\begin{equation}
{\cal
S}_D={\Lambda_{SW}^3\mu^2\over\sqrt2\cdot12\pi}\int_{-1}^{1\over
8\mu^2} dx {x-8\mu^2\over \sqrt{x^2-1}\sqrt{8\mu^2 x -1}},\qquad
{\cal S}={\Lambda_{SW}^3\mu^2\over\sqrt2\cdot12\pi}\int_{-1}^1 dx
{x-8\mu^2\over \sqrt{x^2-1}\sqrt{8\mu^2x -1}}.
\label{opidjcbisse}\end{equation} In terms of $\mu$ the
nonperturbative relation (\ref{v}) reads
\begin{equation}
\mu^4={3\cdot 2^7\pi i\over\Lambda_{SW}^6}\left({\cal F}_0-{{\cal
S}\over 2}{\partial {\cal F}_0\over\partial {\cal S}}\right),
\label{muuu}\end{equation} which is the matrix model analog of the
relation between the $u$--modulus and the Seiberg--Witten
prepotential \cite{rela}.

Introducing manifest duality has several interesting consequences.
For example, one may investigate to what corresponds in matrix
model the strong coupling region of $N=2$ SYM. In particular, the
QFT meaning of the strong coupling expansion of the prepotential
at the points $u=\pm\Lambda_{SW}^2$ is a crucial open question.
While in the weak coupling region the expansion of the SW
prepotential corresponds to a one--loop term and to infinitely
many instanton contributions, no QFT meaning is known for its
expansion at strong coupling. In $N=2$ SYM, this region is
investigated by performing a $S$--duality transformation on the
fields. This corresponds to a Legendre transform of the
prepotential. On the matrix model side one should consider a dual
formulation corresponding to this region. It would be interesting
whether perturbation theory would reproduce also in this region
the $N=2$ SYM theory. One should consider the Legendre transform
\begin{equation}
{\cal F}_{Dg}={\cal F}_g-\sum_{i=1,2}{\cal S}_i{\partial {\cal
F}_g\over\partial {\cal S}_i}, \label{FrankZappa}\end{equation}
where ${\cal F}_g\equiv{\cal F}_g({\cal S}_k,\Lambda,\mu)$, and
\begin{equation}
Z_D= \exp\left(-\sum_{g\geq0}g_{{\cal S}_D}^{2g-2}{\cal
F}_{Dg}\right), \label{expanndual}\end{equation} which should
induce the definition of ${\cal W}_D$
\begin{equation}
Z_D={1\over {\rm vol(G)}}\int d\Phi_D \exp \left(-{1\over g_{{\cal
S}_D}} {\rm tr}\, {\cal
W}_D(\Phi_D)\right).\label{Z_D}\end{equation}

Before concluding, let us note that this approach should be
related with the derivation of the structure of the instanton
moduli space of $N=2$ SYM obtained from the recursion relations
for the instanton contributions to the prepotential \cite{sfere}.
In particular, it was shown how the analogs of the recursive
structure of the Deligne--Knudsen--Mumford compactification of
moduli space of Riemann surfaces and of the Wolpert restriction
phenomenon, essentially determine the structure of the instanton
moduli spaces. These techniques are strictly related to the
geometry of matrix models considered in the framework of Liouville
quantum gravity \cite{LG}. So, it would be interesting to
investigate whether there is a possible link between the matrix
model approach to the $N=2$ SYM and the geometrical approach
considered in \cite{sfere}.

Finally, we note that making duality manifest, which generalizes
to higher rank groups \cite{rangoalto}, may have possible
relations with recent work on matrix models \cite{vari}.

\vspace{.333cm}

\noindent {\bf Acknowledgements}. It is a pleasure to thank A.
Klemm, S. Gukov and M. Mari$\rm \tilde n$o for comments on their
work and G. Bertoldi, G. Bonelli, L. Mazzucato, F. Paccanoni, P.
Pasti, M. Tonin and G. Travaglini for discussions. Work partially
supported by the European Community's Human Potential Programme
under contract HPRN-CT-2000-00131 Quantum Spacetime.

\end{document}